\documentclass[reprint,amsmath,amssymb,aps,superscriptaddress]{revtex4-2}
\usepackage{graphicx}
\usepackage{dcolumn}
\usepackage{bm}

\begin{document}

\title{Chiral-Split Magnon in Altermagnetic MnTe}
\author{Zheyuan Liu}
\affiliation{Institute for Solid State Physics, the University of Tokyo, Kashiwa 277-8581, Japan}
\author{Makoto Ozeki}
\affiliation{Institute for Solid State Physics, the University of Tokyo, Kashiwa 277-8581, Japan}
\author{Shinichiro Asai}
\affiliation{Institute for Solid State Physics, the University of Tokyo, Kashiwa 277-8581, Japan}
\author{Shinichi Itoh}
\affiliation{Institute of Materials Structure Science, High Energy Accelerator Research Organization, Ibaraki 305-0801, Japan}
\affiliation{Materials and Life Science Division, J-PARC Center, Tokai, Ibaraki 319-1195, Japan}
\author{Takatsugu Masuda}
\affiliation{Institute for Solid State Physics, the University of Tokyo, Kashiwa 277-8581, Japan}
\affiliation{Institute of Materials Structure Science, High Energy Accelerator Research Organization, Ibaraki 305-0801, Japan}
\affiliation{Trans-scale Quantum Science Institute, The University of Tokyo, Tokyo 113-0033, Japan}

\date{\today}

\begin{abstract}
Altermagnetism is a newly discovered magnetic class named after the alternating spin polarizations in both real and reciprocal spaces. Like the spin-splitting of electronic bands, the magnon bands in altermagnets are predicted to exhibit alternating chiral splitting. In this work, by performing inelastic neutron scattering on $\alpha$-MnTe, we directly verified the chiral splitting in altermagnetic magnon dispersions. The lifted degeneracy of chirality is further explained by a symmetric-exchange origin. In addition, the $g$-wave magnetism was identified in MnTe. 
% Our results significantly enrich the understanding on altermagnetism from the perspective of magnon bands, motivating further exploration in this novel magnetic phase.
\end{abstract}

\maketitle

Altermagnets have been recently categorized as a third elementary type of collinear magnetic phases in addition to the conventional ferromagnets and antiferromagnets~\cite{PhysRevX.12.040501,PhysRevX.12.031042}. The distinct spin symmetries in altermagnets simultaneously keep the compensated spin arrangement and break the time-reversal symmetry (TRS). Consequently, electronic bands in altermagnets are spin-splitting even in the absence of the relativistic spin-orbit coupling. The unconventional spin-splitting has been predicted to be significant owing to the exchange origin~\cite{doi:10.7566/JPSJ.88.123702,PhysRevB.102.144441,PhysRevB.102.014422,PhysRevMaterials.5.014409,C5CP07806G,PhysRevB.99.184432,doi:10.1021/acs.jpclett.1c00282,doi:10.1073/pnas.2108924118,GUO2023100991,PhysRevB.105.054402,PhysRevB.109.094425}, and has been subsequently observed in altermagnetic candidates MnTe~\cite{krempasky2024altermagnetic,PhysRevLett.132.036702,PhysRevB.109.115102}, RuO$_2$~\cite{doi:10.1126/sciadv.adj4883}, and CrSb~\cite{reimers2024direct} by angle-resolved photoemission spectroscopy (ARPES). Besides the band splitting, altermagnets exhibit various exotic quantum phenomena related to the TRS breaking nature, such as the anomalous Hall effect~\cite{doi:10.1126/sciadv.aaz8809,vsmejkal2022anomalous,feng2022anomalous}, charge-spin current conversion~\cite{PhysRevLett.126.127701,naka2019spin,shao2021spin,ma2021multifunctional,PhysRevB.103.125114,bose2022tilted}, spin splitter torque~\cite{PhysRevLett.129.137201,PhysRevLett.128.197202}, magnetic circular dichroism~\cite{PhysRevLett.132.176701}, and piezomagnetic effect~\cite{ma2021multifunctional,PhysRevMaterials.8.L041402}. These remarkable findings establish a firm foundation for the altermagnetic spintronics functionalities~\cite{PhysRevX.12.040501}.

\begin{figure}[htbp] 
\includegraphics[width=\linewidth]{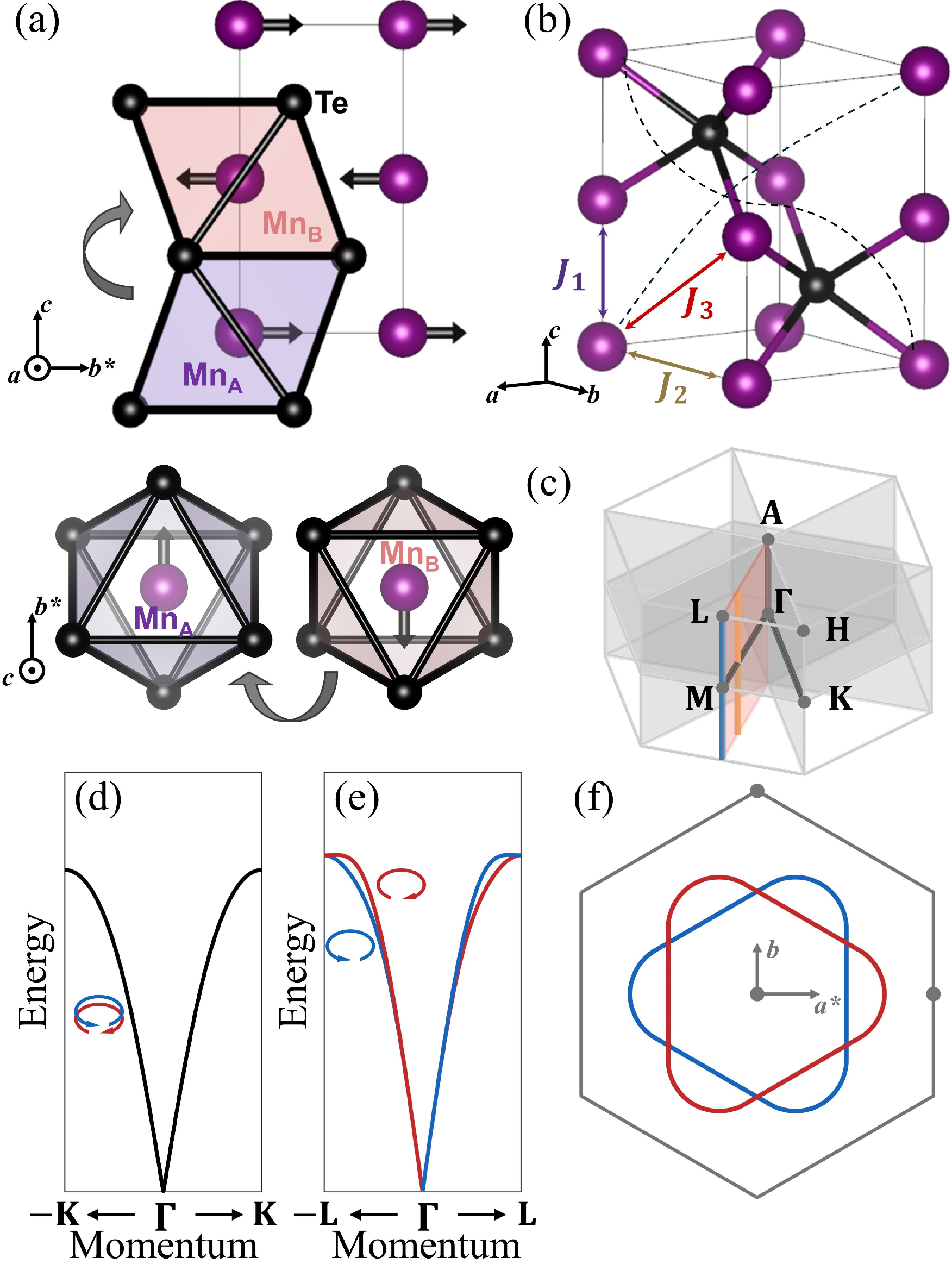}
\caption{\label{Fig_1}
(a) Lattice and spin structure of MnTe. The two opposite-spin sublattices, Mn$_\text{A}$ and Mn$_\text{B}$, are connected by a real-space mirror or sixfold rotation symmetry. (b) Exchange paths between Mn$^{2+}$ spins. The dashed curves highlight the altermagnetic superexchange paths. (c) Schematics of the MnTe BZ with four spin-degenerate NPs (in gray). The $\Gamma$MLA-plane is indicated in red. Schematics of altermagnetic magnon dispersion (d) in the NPs along $\Gamma-\text{K}$ and (e) off the NPs along $\Gamma-\text{L}$. (f) The constant-energy contour of the altermagnetic magnon dispersions in a $c^{*}$-plane off the NPs and BZ boundaries.}
\end{figure}

Similar to the spin-splitting of electronic bands, the altermagnetic magnons are theoretically predicted to exhibit chiral splitting within the consideration of symmetric exchange interactions~\cite{PhysRevLett.131.256703,PhysRevB.108.L100402}. The spin current in ferromagnets carried by the chiral magnons is only operational at slow GHz rates due to the quadratic magnon dispersion. By contrast, the linear dispersive chiral magnons in altermagnets reach the THz frequencies, unveiling the potential for the stray-field-free ultra-fast spintronics~\cite{PhysRevX.12.040501,PhysRevLett.131.256703}. In this letter, we performed inelastic neutron scattering (INS) on altermagnetic MnTe single crystals. The non-degenerated magnon dispersions are unambiguously verified in the observed spectra. The obtained spin Hamiltonian demonstrates that the alternating superexchange interactions lift the magnon degeneracy for both energy and chirality. Our result provides the evidence for altermagnetic magnon chiral splitting and highlights the exchange origin of altermagnetic nature.
%bulk property

\begin{figure*}
\includegraphics[width=\linewidth]{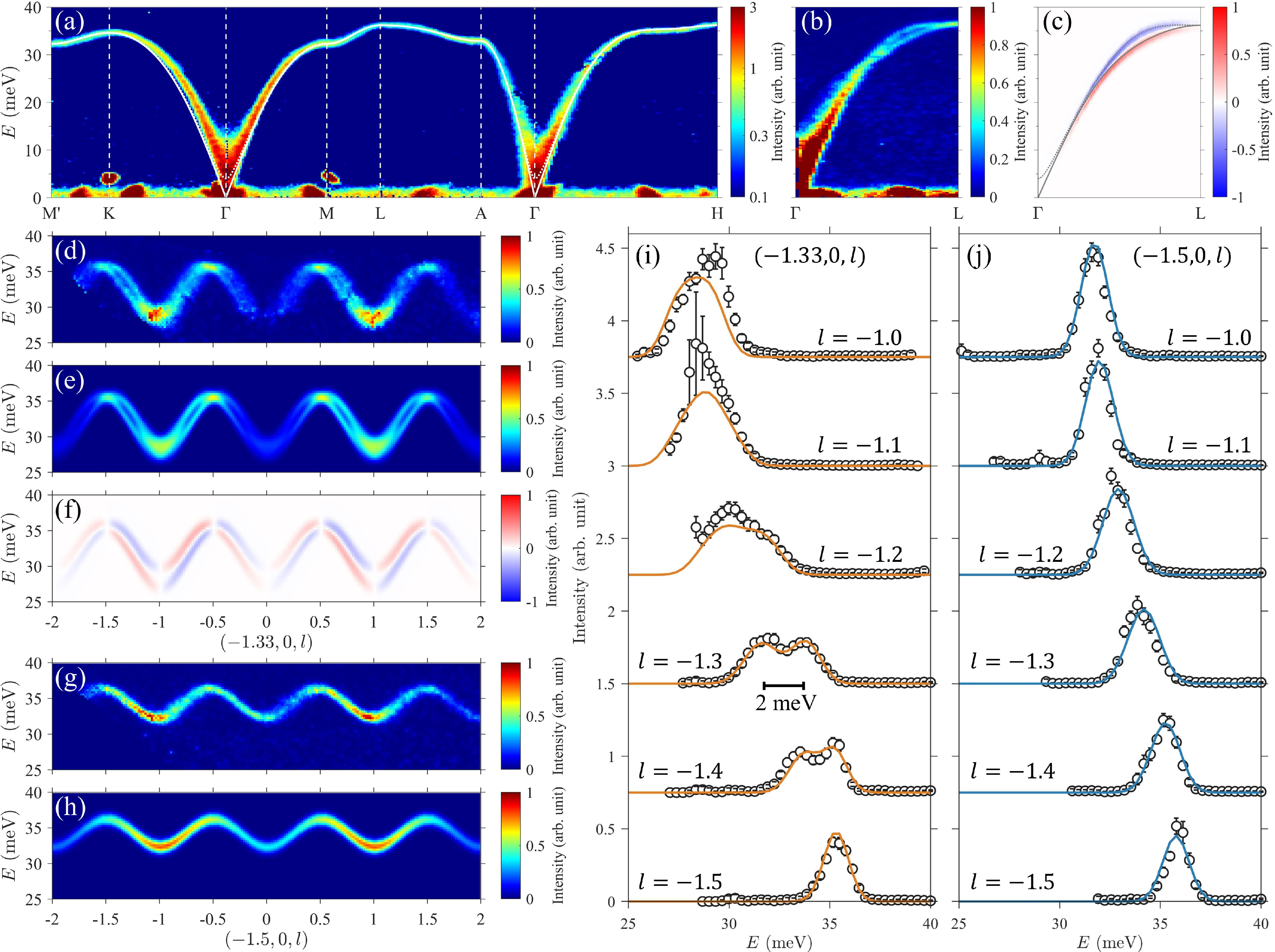}
\caption{\label{Fig_2} INS spectra with $E_{i}$=50.8 meV at $T$=10 K and LSWT calculation results. (a) INS spectra along high symmetric $Q$-paths inside the spin-degenerate NPs. The colorbar is in logarithm scale. The white solid and dotted curves show two calculated magnon modes. (b) INS spectrum and (c) calculated neutron chiral factor along $\Gamma-\text{L}$ off the spin-degenerate NPs. The gray solid and dotted curves show two calculated magnon modes. (d) INS spectra, (e) calculated neutron structure factor, and (f) calculated neutron chiral factor along $(-1.33,0,l)$ off the NPs. (g) INS spectra and (h) calculated neutron structure factor along $(-1.5,0,l)$ in the BZ boundary. (i), (j) 1-D constant-$Q$ cuts of the INS spectra and calculated neutron structure factors along $(-1.33,0,l)$ and $(-1.5,0,l)$, respectively. The black circles are experimental data and the solid curves are calculations. The $(-1.33,0,l)$ and $(-1.5,0,l)$ directions correspond to the orange and blue paths shown in Fig.~\ref{Fig_1}(c), respectively.} 
\end{figure*}

The studied material, $\alpha$-MnTe, hosts a centrosymmetric hexagonal lattice with a space group $P\text{6}_\text{3}/mmc$, as shown in Figs.~\ref{Fig_1}(a) and~\ref{Fig_1}(b). Below N\'{e}el temperature $T_{N}=307$ K, an $A$-type antiferromagnetic (AFM) ground state was realized where the spins $S=5/2$ of Mn$^{2+}$ ions are ferromagnetically orientated along the crystallographic $b^{*}$-axis in the $ab$-plane and antiferromagnetically arranged between adjacent layers~\cite{EFREMDSA2005267,https://doi.org/10.1002/pssc.200460669,PhysRevB.96.214418}. The opposite-spin sublattices, Mn$_\text{A}$ and Mn$_\text{B}$, are connected by a mirror operation or sixfold rotation instead of any translation or inversion, which meets the classification of altermagnets, as shown in Fig.~\ref{Fig_1}(a). The mirror and sixfold rotation symmetries protect the degeneracy for electronic bands in four nodal planes (NPs)~\cite{krempasky2024altermagnetic,PhysRevLett.132.036702}, as shown in Fig.~\ref{Fig_1}(c), in addition to the Brillouin zone (BZ) boundaries. The bulk-sensitive X-ray ARPES on MnTe revealed a half-eV-scale spin-splitting in the electronic bands off the NPs~\cite{krempasky2024altermagnetic}, suggesting a significant altermagnetic effect in MnTe.

Based on the spin symmetry, the altermagnetic superexchange interactions in MnTe are analyzed to be the tenth and eleventh nearest neighbor interactions, $J_{10}$ and $J_{11}$, which alternatively connect the same-spin sublattices, as illustrated by the dashed curves in Fig.~\ref{Fig_1}(b). Although the distanced interactions $J_{10}$ and $J_{11}$ are inevitably weak, they are expected to differ in value or even in sign because of the different exchange paths through the Te orbitals. Similar to the electronic bands, the degeneracy of the chiral magnons is found to be protected by the symmetries between the sublattices as well. Inside the NPs and BZ boundaries, the altermagnetic magnon dispersion is identical to the conventional AFM magnon dispersion, as illustrated by the dispersion along $\Gamma-\text{K}$ shown in Fig.~\ref{Fig_1}(d). However, a finite difference between $J_{10}$ and $J_{11}$ lifts the magnon degeneracy off the NPs and BZ boundaries, while the altermagnetic magnons still maintain the AFM-like linear dispersion near the $\Gamma$-point, as illustrated by the dispersions along $\Gamma-\text{L}$ shown in Fig.~\ref{Fig_1}(e). The constant-energy contour of the non-degenerated dispersions in the $c^{*}$-planes shows a $g$-wave harmonic symmetry in MnTe, as shown in Fig.~\ref{Fig_1}(f). Note that these non-degenerated dispersions are alternating chiral-split as well. Previous INS experiments on MnTe limited by a relatively relaxed energy resolution only probed the first, second and third nearest neighbor exchange interactions (denoted as $J_1$, $J_2$, and $J_3$)~\cite{PhysRevB.73.104403}, and would miss the splitting feature of the magnon dispersions. 

\begin{figure}[htbp] 
\includegraphics[width=\linewidth]{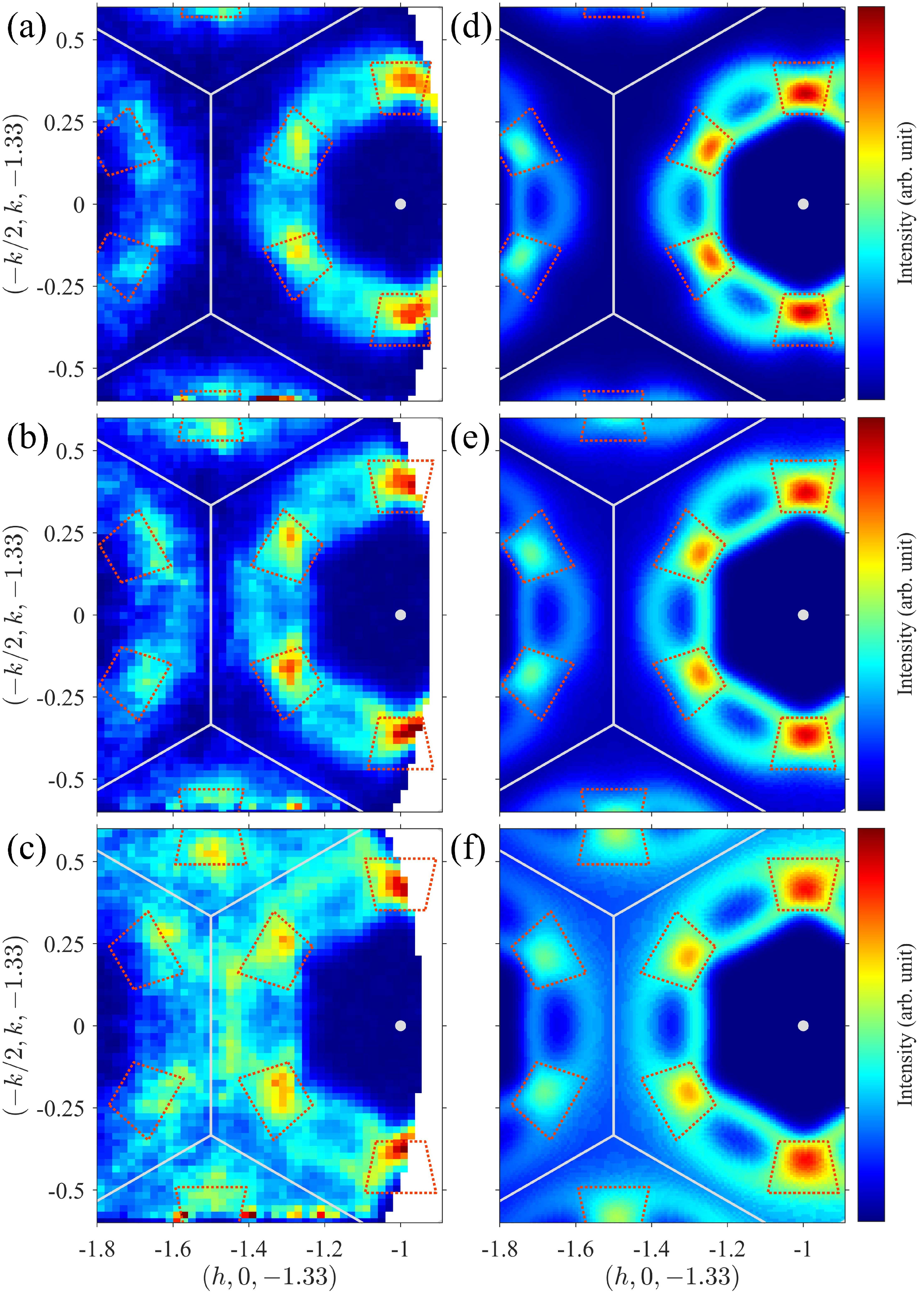}
\caption{\label{Fig_3}
Constant-energy slices in the $(h,k,-1.33)$-plane of (a) $E=33$ meV, (b) $E=33.5$ meV, and (c) $E=34$ meV. (d), (e), (f) LSWT calculations corresponding to (a), (b), and (c). The red dotted trapezoids highlight the six-fold nodal points of the $g$-wave contour of the altermagnetic magnon dispersion. The BZs are shown by the solid white lines.}
\end{figure}

To perform INS, large MnTe single crystals were synthesized by the self-flux method~\cite{MATEIKA1972698}. 
%Pure Mn and Te with molar composition Mn$_{0.436}$Te$_{0.564}$ were placed in an alumina crucible and sealed in a quartz under vacuum. The excess Te was used for lowing the crystallization temperature of MnTe below the structural transition temperature at 1026 $^{\circ}$C. The remaining melt was removed at 760 $^{\circ}$C to avoid peritectic formation of MnTe$_2$. 
Five crystals with total mass of 4.2 g were co-aligned with a scattering plane spanned by $(h,0,0)$ and $(0,0,l)$. The INS experiments were carried on the High Resolution Chopper (HRC) spectrometer installed at BL-12 in MLF, J-PARC~\cite{ITOH201190,ITOH201976,Kawana_2018}. A Gifford-McMahon type cryostat was used for controlling the temperature. The frequency of Fermi chopper was 200 Hz. 
The incident energies, $E_i$s, of 50.8 meV and 11.4 meV were used for the measurements at $T = 10$ K.
%The incident energy, $E_i$, of 50.8 meV was used for the measurements at $T = 10$ K. 
HRC has vertical detector banks, thus accesses the momentum ($Q$) space out of the scattering plane. 
% For the data shown in Figs. 2(a) and 2(b), the reciprocal space is folded along all the $a^{*}$, $c^{*}$, and $b$-axes. For the other data in this letter, the reciprocal space is not folded.

Well-defined magnon excitations were observed by the INS experiments with $E_i=50.8$ meV, as shown in Fig.~\ref{Fig_2}. The spectrum along high symmetric $Q$-paths, M$'(-\frac{1}{2},\frac{1}{2},1)$-K$(-\frac{2}{3},\frac{1}{3},1)$-$\Gamma(-1,0,1)$-M$(-\frac{3}{2},0,1)$-L$(-\frac{3}{2},0,\frac{3}{2})$-A$(-1,0,\frac{3}{2})$-$\Gamma(-1,0,1)$-H$(-\frac{1}{2},\frac{1}{2},\frac{3}{2})$, which are all inside the spin-degenerate NPs, is shown in Fig.~\ref{Fig_2}(a). It presents a typical AFM-like magnon dispersions with a single-ion easy-plane anisotropy. The spots around $E=5$ meV at K and M points are considered as artifacts, since they are absent in the spectrum with $E_i=11.4$ meV. In contrast to the spectrum in the NPs, the dispersions off the NPs and BZ boundaries are distinct from the AFM magnon dispersions. Spectrum along $\Gamma(-1,0,1)$-L$(-\frac{1}{2},0,\frac{3}{2})$ in Fig.~\ref{Fig_2}(b) shows a magnon splitting above $E\sim 30$ meV, while the dispersions keep the linear feature in the lower energy range. This splitting gradually smears when approaching $\Gamma$ or L in the spin-degenerate NP. The split dispersions were clearly observed along the $(-1.33,0,l)$ direction, as shown in Fig.~\ref{Fig_2}(d). The two non-degenerated magnon modes alternatively propagate along the $c^{*}$-axis, forming nodal points at integer and half-integer $l$, which implies the mirror symmetry of the Mn$^{2+}$ sublattices. For comparison, we present a similar dispersion along $(-1.5,0,l)$ in the BZ boundary, as shown in Fig.~\ref{Fig_2}(g). The one-dimensional (1-D) constant-$Q$ cuts of the split dispersions along $(-1.33,0,l)$ in Fig.~\ref{Fig_2}(i) demonstrate a double-peak splitting of ~2 meV in most of the intermediate $l$ range ($-1<l<-1.5$). In contrast, at $l=-1$, $-1.5$ and every cut along $(-1.5,0,l)$ in Fig.~\ref{Fig_2}(j), the excitation is a sharp single peak.

The $g$-wave harmonic symmetry of the altermagnetic MnTe which originates from the improper sixfold rotation between the Mn$^{2+}$ sublattices~\cite{PhysRevX.12.040501,PhysRevX.12.031042} was unambiguously verified in the constant-$E$ slice in the off-nodal $(h,k,-1.33)$-plane, as shown in Figs.~\ref{Fig_3}(a)-~\ref{Fig_3}(c). The sixfold high intensity nodal points were clearly observed at $E=33$ meV to $34$ meV. The split contour of the dispersion shows a typical intertwined $g$-wave pattern, again relating the unconventional splitting to the spin symmetry.

To identify the spin Hamiltonian, calculations were performed based on the linear spin-wave theory (LSWT) by using SpinW package~\cite{toth2015linear}. Because the concept of altermagnetism is within the category of symmetric exchange interaction, we consider a Heisenberg spin Hamiltonian with a single-ion easy-plane anisotropy, represented by
\begin{equation}
{\mathcal H} = \sum_{\left\langle i,j\right\rangle }{J_{ij}\bm{S}_{i}\cdot \bm{S}_{j}}+\sum_{i}{D(S^{z}_{i})^2}, 
\label{Eq_1}
\end{equation}
where $i$ denotes the index of Mn$^{2+}$ spin and $\bm{S}_{i}$ is the spin operator. $J_{ij}$ is the exchange interactions between $i$th and $j$th spins. The sum is taken over pairs of spins. $D$ is the anisotropy constant, and $S^{z}$ is the $z$ component of the spin operator, where $x$, $y$, and $z$ are the crystallographic $a^{*}$, $b$, and $c^{*}$ axes, respectively. 

During the simulation, we verified that neither the dominant exchange interactions, $J_1$, $J_2$, and $J_3$, nor the further forth to ninth nearest neighbor exchange interactions, could explain the split dispersions in Figs.~\ref{Fig_2}(b) and \ref{Fig_2}(d). The unconventional magnon splitting are irrelevant to relativistic effect as well; Dzyaloshinskii-Moriya interaction is inactive in the centrosymmetric lattice of MnTe, and the single-ion easy-plane anisotropy only lifts the degeneracy near the $\Gamma$-point. Thus, the altermagnetic exchange interactions, $J_{10}$ and $J_{11}$, are crucial in the spin Hamiltonian.

Inside the NPs, $J_1$, $J_2$, $J_3$, $J_{10}+J_{11}$ and $D$ are the parameters for the dispersion formula. By fitting to the spectra in the NPs by the weighted least squares method, the best solution gives $J_1=3.99(3)$ meV, $J_2=-0.120(2)$ meV, $J_3=0.472(3)$ meV, $J_{10}+J_{11}=0.0931(8)$ meV and $D=0.0482(5)$ meV. The calculated dispersions are over plotted in Fig.~\ref{Fig_2}(a), suggesting the high consistency. Then, by fitting to the spectra off the NPs, we obtained $J_{10}=0.0681(7)$ meV and $J_{11}=-0.0221(2)$ meV. The calculated neutron structure factors after being convoluted by the instrumental energy resolution are shown in Figs.~\ref{Fig_2}(e) and~\ref{Fig_2}(h). Both the single- and double-peak magnon excitations in Figs.~\ref{Fig_2}(i) and~\ref{Fig_2}(j) are well explained by the obtained spin Hamiltonian. The $g$-wave patterns in the constant-$E$ slices are reproduced by the calculations as well, as shown in Figs.~\ref{Fig_3}(d)-~\ref{Fig_3}(f).

To further verify the chirality of the split magnon modes, we calculated the neutron chiral factor based on the obtained spin Hamiltonian by using LSWT. The chiral factor is represented by
\begin{equation}
M_{\mathrm{ch}}=i(\langle M_{Y}M_{Z}^\dagger\rangle-\langle M_{Z}M_{Y}^\dagger\rangle), 
\label{Eq_2}
\end{equation}
where $X$ is parallel to $\bm{Q}$, $Y$ is orthogonal to $X$ in the scattering plane, and $Z$ is orthogonal to the scattering plane. $\langle M_{\alpha}M_{\alpha}^\dagger\rangle\ (\alpha=Y,Z)$ is the Fourier transform of the spin-spin correlation function. The calculated $M_{\mathrm{ch}}$ along $\Gamma-\text{L}$ is shown in Fig.~\ref{Fig_2}(c). The magnon modes split by the altermagnetic exchange interactions exhibit $M_{\mathrm{ch}}$ with opposite sign, indicating an opposite chirality. Note that $M_{\mathrm{ch}}$ is zero at the $\Gamma$-point. Though the easy-plane anisotropy lifts the two-fold degeneracy of magnon at the $\Gamma$-point, the chiral splitting is prohibited here because the TRS is still preserved. Similarly, we plotted $M_{\mathrm{ch}}$ along $(-1.33,0,l)$, as shown in Fig.~\ref{Fig_2}(f). The alternatively propagating magnon dispersions are alternating chiral-split as well. By contrast, the magnon chirality is completely degenerated and the $M_{\mathrm{ch}}$ is zero along $(-1.5,0,l)$. These evidence the chiral magnons in a magnet with collinear compensated spin arrangement, in addition to the chiral magnon observed in ferromagnet~\cite{PhysRevB.105.L180408}, ferrimagnet~\cite{PhysRevLett.125.027201}, and non-collinear $120^{\circ}$ antiferromagnet~\cite{PhysRevLett.106.207201}.

The identified altermagnetic exchange interactions are different in value and sign, accentuating the distinct spin symmetry of altermagnetism. Though weaker than the dominant interactions in the system by order of magnitude, they bring significant outcomes. As an example, they enable the linear dispersive chiral magnons in altermagnets near the $\Gamma$-point, removing the limitation of ferromagnetic magnon spintronics~\cite{PhysRevX.12.040501}. Also, in addition to the $s$-wave ferromagnetism, the altermagnetic exchange interactions lead to $d,g,i$-wave magnetic phases, complementing the long missing counterpart of unconventional superconductivity in magnetism~\cite{PhysRevX.12.040501,PhysRevB.108.184505}.

In conclusion, by performing INS measurements, we verified the chiral splitting of magnon in the altermagnet and confirmed the $g$-wave magnetism in MnTe. The giant splitting of magnon dispersions reaches ~2 meV and is well explained by a pair of alternating exchange interactions. By providing the evidence of altermagnetism from the perspective of spin excitation, this work established a firm foundation for future explorations in this new magnetic phase.

\begin{acknowledgements}
We are grateful to D. Kawana, T. Asami, and R. Sugiura for supporting us in the neutron scattering experiment at HRC. The neutron experiment using HRC at the Materials and Life Science Experimental Facility of the J-PARC was performed under a user program (Proposal No. 2024S01). Z. Liu was supported by the Japan Society for the Promotion of Science through the Leading Graduate Schools (MERIT). This project was supported by JSPS KAKENHI Grant Numbers 21H04441. 
\end{acknowledgements}

%\bibliography{Ref.bib}
%

\end{document}